# Degradation of organic compounds and production of activated species in Dielectric Barrier Discharges and Glidarc reactors.

J. M. Cormier, O. Aubry, A. Khacef

GREMI - Polytech'Orleans, 14 rue d'Issoudun, BP 6744, 45067 Orléans Cedex 2, France

**Abstract**

Major sterilization mechanisms are related to atoms and radicals, charged particles, excited molecules, ozone, and UV radiation. The ROS (Reactive Oxygen Species) are well known as evildoers. These species are easily created in ambient air and water and they live long enough to reach the cell and attack the organic matter. Test molecules conversion in dry and wet air is studied using Dielectric Barrier Discharge (DBD) and Gliding Arc Reactors (GAR). The effects of temperature and energy deposition into the media on the active species production and then on the organic compounds degradation are presented for two non thermal plasma reactors: DBD and GAR. Main production species investigated are OH, $O_3$, NOx, CO and $C_xH_yO_z$ by-products. It is shown from experiment analysis that the reactive species production is quite different from one reactor to another. GAR and pulsed DBD are two chemical processing ways in which the temperature of heavy species in ionized gas is determinant. By reviewing the species production obtained from both reactors, a discussion is open about plasma decontamination.

## 1. Introduction

Non-thermal atmospheric pressure plasma processing is one of the most effective technology for removal pollutants (nitrogen oxides NOx, sulfur oxides, volatile organic compounds VOCs) from flue gas [1-11] and syngas gas production from hydrocarbons and alcohols [12-15] at relatively low energetic costs. NTP that has a low gas temperature and a high electron temperature can be produced by a variety of electrical discharge methods (pulsed corona discharge, barrier discharge, and dc discharge, gliding arc) [16-22] or electron beam irradiation [1, 23]. The energy supplied into the discharge is used preferentially to create free electrons, which are then used to produce ionization and excitation of the gas mixture components. Gas phase radicals such as hydroxyl (OH), hydroperoxyl (HO$_2$) and oxygen atoms (O) were generated and were consumed in chemical reactions, part of them promoting the desired conversion of pollutants. In this last case, high electron energy is observed although the gas remaining at room temperature. In the case of non thermal arcs operating at low current intensity (I < 2A), electron temperature is generally higher than 8000 K and gas temperature smaller than 6000 K. Non thermal behavior is quite different in this case because of a sufficient gas temperature able to ignite chemical reactions. Mechanisms in high voltage transient discharges (DBD, Corona) and non thermal arc discharges (Glidarc) lead the production of different species. Our main interest is the nature of these species as well as their properties for sterilisation, decontamination and medical applications.

## 2. The gliding discharge plasma reactor "Glidarc"

Gliding arc discharges had been presented previously [20-22]. The electrical system consists of two arcing horns and a DC high voltage power supply (fig. 1).

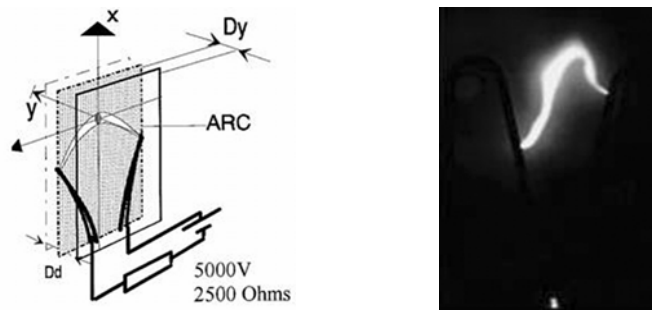

**Fig. 1.** Schematic of gliding arc reactor and photography of the discharge

Through the inlet nozzle, the reacting gas flows and blows the arc. The vessel is built with two electrodes made of copper tubes. These tubes have a diameter of 8 mm and a length of 20 cm. Electrodes is gripped between two panes of glass. These panes are spaced of 1.2 cm and insulated the area between the electrodes. The arc characteristics (gas flow rate, electric field strength, power density), gas temperature, plasma radii-conduction and apparent, are given in table 1. Direct current intensity is 0.9 and 1.7 A. Spectroscopic measurements show that electron temperature is in the range 10 000-12 000 K.

| Flow (Nm$^3$/h) | E (V/m) | E I (W/m) | T (K) | Conduction radius (mm) | Apparent radius (mm) |
|---|---|---|---|---|---|
| 3 | 7700 | 7000 | 5200 | 1,6 | 1,4 |
| 8 | 14500 | 24700 | 6200 | 0.8 | 0.7 |

**Table 1.** Arc characteristics

According to Cavvadias [4], the mechanisms for NO production at low temperature (T < 4000 K) (ions are not taken into account) are given by the following chemical reactions

$$O_2 + M \underset{k_2}{\overset{k_1}{\Longleftrightarrow}} 2\,O + M \qquad O_2 + N \underset{k_8}{\overset{k_7}{\Longleftrightarrow}} NO + O$$

$$N_2 + M \underset{k_4}{\overset{k_3}{\Longleftrightarrow}} 2\,N + M \qquad N_2 + O_2 \underset{k_{10}}{\overset{k_9}{\Longleftrightarrow}} NO + O$$

$$O + N_2 \underset{k_6}{\overset{k_5}{\Longleftrightarrow}} NO + N \qquad NO + M \underset{k_{12}}{\overset{k_{11}}{\Longleftrightarrow}} N + O + M$$

Solutions of the set of the coupled differential equations are given using "Chemical Workbench" code and kinetic constants were taken from NIST data base.

Let us consider a one dimensional air flow device with a hot region on which temperature is increasing from A to B and decreasing from B to C. The temperature profile is supposed to be calculated as a function of the distance: x (fig. 2). The air molecules are flowing from A to D at a velocity v. A is corresponding to the input and D to the output. From A to D the air flow is submitted to heating and cooling. The transition time from hot region crossing is: $t_C - t_A = (x_C - x_A)/v$.

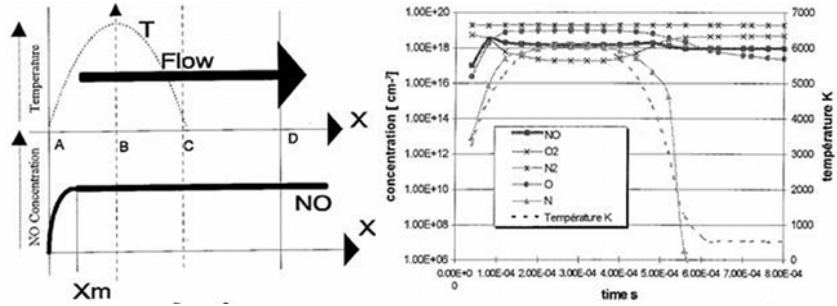

**Fig. 2.** Schematic picture of NO formation and calculated concentration species for an axial temperature of 5000 K.

The NO concentration is increasing as the molecules are flowing in the heated zone and a maximum value of NO concentration is reached at a distance $X_m$ which corresponds to a flowing duration time of $\delta t = t_A - t_{Xm}$. In a general case, $X_m$ does not correspond to the maximum temperature because NO production is decreasing at high temperature. Solving kinetic equations of the above chemical reactions, NO concentration at the output of the device can be calculated. Simulations were carried out using temperature profiles and gas velocities obtained from the plasma string model. Example of results is given on fig. 2. Calculation leads to 2700 ppm and 500 ppm of NO for low rate of 3 Nm³/h and 8 Nm³/h, respectively. The measured NO concentrations are 3000 ppm and 500 ppm respectively.

Analysis of NO production shows that NO production chemistry can be described by a quasi-thermal process. NO oxidation leads to $NO_2$ production. Measured $N_2O$ concentration remains at low values (<1%) in all cases. Mains chemical effects after reaction with water lead to acids production ($HNO_2$, $HNO_3$).

## 3. Rotating discharge plasma reactor

The "rotating discharge" reactor (fig. 3) operating at atmospheric pressure was powered with a 50 Hz high voltage step-up transformer, 220 V/10 kV, with leakage fluxes. The effect of leakage fluxes determines a reactance that produces a constant RMS value of the discharge current (100-200 mA). The reactor consists of a quartz tube in which a conical central electrode and a cylindrical external are laid out. The gas is injected into the reactor transversely with the axis of the system. The voltage applied to the electrodes can reach several tens of KV. The discharges are produced between the electrodes at the place where those are closest.

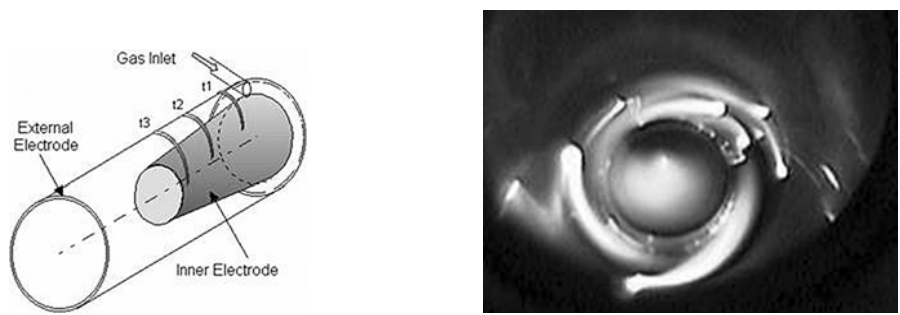

**Fig. 3.** Rotating discharge reactor and photography of the rotating discharge

Experiments with "Rotarc" were performed at room temperature [10] using the following mixture: $O_2$ (10%) - NO (500 ppm) - $C_3H_6$ (500 ppm) and $N_2$ as balance at 1 atm. Output species are shown on fig. 4.

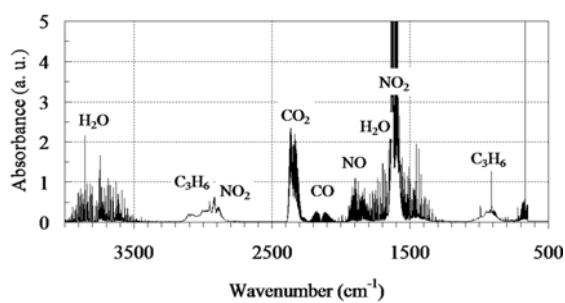

**Fig. 4.** Typical FTIR spectrum ($O_2$ (10%)-NO (500 ppm) - $C_3H_6$ (500 ppm)-$N_2$ mixture)

5Main detected products are CO, $CO_2$, NO, $NO_2$, and $H_2O$. Output species concentrations obtained after plasma processing were compared to those without plasma. Results are shown on the table 2.

| Species | Plasma OFF | Plasma ON |
|---------|------------|-----------|
| NO | (539 ± 27) ppm | (1838 ± 92) ppm |
| $NO_2$ | (27 ± 5) ppm | (362 ± 18) ppm |
| NOx | (566±28) ppm | (2200 ± 110) ppm |
| CO | (10 ± 5) ppm | (98 ± 5) ppm |
| $CO_2$ | (4 ± 2) ppm | (117 ± 5) ppm |
| $C_3H_6$ | (508 ± 10) ppm | (438 ± 10) ppm |

**Table 2.** Output concentration of mains products

Results show the ability of rotating discharge for high values of NOx concentrations production in air. NOx repartition is NO (84%) and $NO_2$ (16%). NOx production in rotarc and Gliding discharge is similar and very effective because of temperature of the plasma gas column. As the 210 ppm of input atoms of carbon lead to CO (98 ppm) and $CO_2$ (117 ppm) carbon balance is satisfied. These results show that no other species are produced in the limits of the detection capacity of the FTIR. Physics and chemistry in rotating discharge and gliding arc reactor are similar.

## 4. Stationary non thermal plasma "Statarc"

The plasma was created between two carbon electrodes with few mm gap spacing (fig. 5) at atmospheric pressure. One may underline a typical feature of the plasma which is the difference between electron and gas temperature.

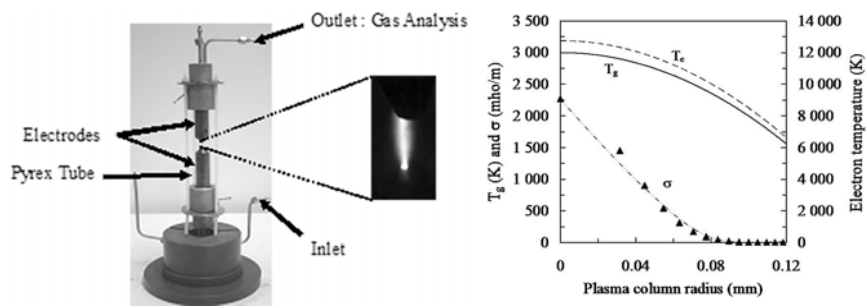

**Fig. 5.** "Statarc" reactor photograph and plasma characteristics plasma column (case of $N_2$)

The device has been mainly tested for the conversion of methane in various mixtures (Steam and air). Typical results of conversion are shown in fig. 6.

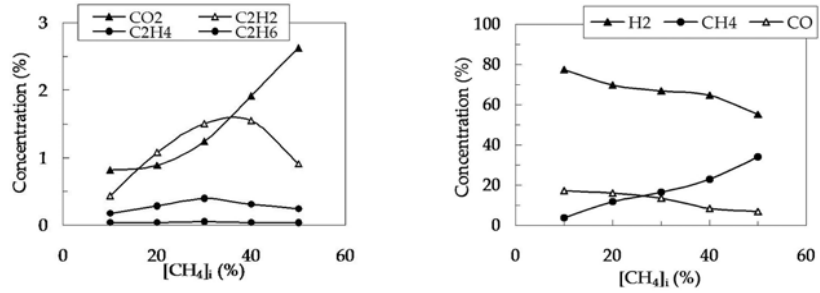

**Fig. 6.** Exhaust gas composition *vs* inlet CH$_4$ concentration in a steam reforming plasma reactor.

The plasma was modeled by FEMLAB software by using a simplified kinetic model. Plasma is supposed to be described by the following energy equation in which transport properties are polynomial fitted as function of plasma temperature.

$$\rho C_p \frac{\partial T}{\partial t} + \vec{\nabla}\left(-\lambda \vec{\nabla} T + \rho C_p T \vec{u}_s\right) = \sigma E^2$$

Kinetic is described by the following simplified equation in which rate constant is a function of temperature.

$$\vec{\nabla}\left(-D\vec{\nabla} y + y\vec{u}_s\right) = k\left([CH_4]_0 - y\right)\left([H_2O]_0 - y\right)$$

A Kinetic rate expression ($k$ (T)) deduced from the fits obtained with results of GRI-3 mechanism is used. A picture of a discharge and CH$_4$ conversion map are shown on fig. 7.

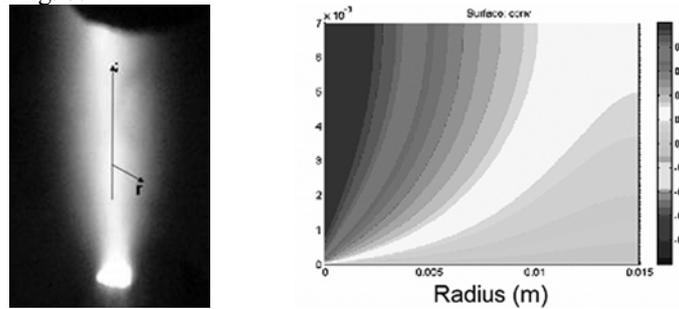

**Fig. 7.** Photography of the plasma column and methane map conversion (calculated conversion rate: 40%, measured conversion rate: 48%)

Experimental and calculated conversions are compared. Taking into account roughly approximations in plasma modeling one may conclude in a useful first description given. "Statarc" studies show that a quasi-thermal analysis leads to results in good agreement with experiments. "Statarc" can be considered as elemen-tary reactor in which main phenomena are similar that the ones involved in "Glidarc" and "Rotarc".

## 5. Non thermal arc conclusion

"Glidarc", "Rotarc", and "Statarc" reactors involve similar physical and chemical processes, and basically non equilibrium plasma: electron temperature is higher than gas temperature. First chemical approach can be performed using simplified kinetics and plasma fluid model.

- Main species produced in air and $CH_4$-$H_2O$ mixtures are NOx, CO, $CO_2$, $C_2H_2$, $C_2H_6$ and $H_2$.
- Oxidized high hydrocarbon species are observed at very low concentrations and not significant for such a plasma treatment.
- Others species can be produced in post discharge regions by reaction with added products.
- Because of output temperature higher than 200°C no ozone production is available from these devices.

## 6. Non-thermal plasma study in VOCs' treatment: propane conversion in a pulsed DBD reactor

Non-thermal atmospheric pressure plasma processing is being actively studied for removal of volatile organic compounds (VOCs') diluted in air (25-30). In spite of the well-established implementations, experimental and fundamental knowledge of chemical processes need to be improved, especially products analysis and mass balance.

Propane ($C_3H_8$) decomposition was investigated using a pulsed high voltage Dielectric Barrier Discharges (DBD) reactor in a wire to cylinder configuration. At room temperature, the propane removal efficiency and by-products production depends strongly on input energy density ($E_d$). Main carbon species produced are CO and $CO_2$. Others $C_xH_yO_z$ (formaldehyde, formic acid, …) species can be produced for the lowest energy density studied.

A temperature effect (up to 800 K) on propane conversion and species production was studied. Main results show that an increase of the temperature leads to an increase of the propane consumption (fig. 8). High propane conversion rates are obtained for the lowest $E_d$ at high gas temperature. As example, for $E_d = 100$ J.L$^{-1}$, $C_3H_8$ is entirely consumed at 800 K whereas only 50% are converted at 450 K.

Plasma plays a major role to initiate propane decomposition. Without plasma, no reactions are observed even for the highest temperature investigated. In terms of energetic cost for VOCs conversion, the gas temperature should be taken into account. Thermal energy, $E_{th}$, can be calculated. In comparison to 300 K, $E_{th}$ contribution is about 120 and 220 J.L$^{-1}$ for 450 and 800 K, respectively. Temperature variation leads to a dramatic effect on the species produced. Increasing the temperature promotes $CO_2$ which becomes the main specie in the outlet gas (fig. 9).

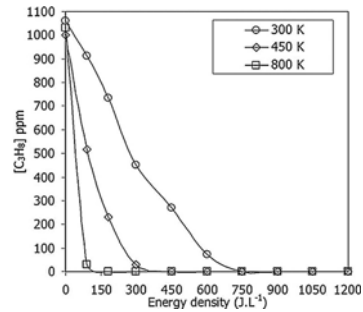

**Fig. 8.** Outlet $C_3H_8$ concentration *vs* $E_d$ for 300, 450 and 800 K (Charging voltage=5 kV).

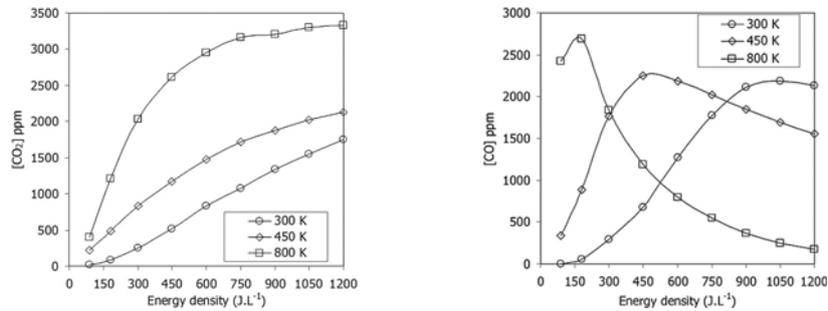

**Fig. 9.** Outlet CO and $CO_2$ concentration *vs* input energy density for 300, 450 and 800 K.

In Fig. 10 are reported measured carbon concentrations ([MC]) from non-consumed $C_3H_8$ and CO and $CO_2$ produced ([MC] = 3[$C_3H_8$] + [CO] + [$CO_2$]). Carbon deficit increases when gas temperature decreases. Input energy density range where carbon balance is not complete depends strongly on gas temperature. At room temperature, carbon deficit (ΔC) is observed in the range 50-900 $J.L^{-1}$ whereas at 450 K, the ΔC range is reduced to 50-400 $J.L^{-1}$. This means that the number or the concentration of non-measured carbon species increase. These species such as formaldehyde, acetaldehyde, formic acid, and others $C_xH_yO_z$ are observed experimentally but are not quantified in that study.

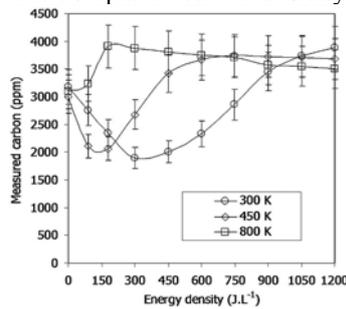

**Fig. 10.** Measured carbon *vs* input energy density for different gas temperature

Ozone ($O_3$) production seems to be an important parameter to explain measured carbon profiles. Maximum carbon deficit measured can be linked to the maximum ozone concentration. In fig. 11 are reported ozone concentrations as functions of input energy density for gas temperature up to 800 K. When $O_3$ concentration decreases, carbon deficit decreases and corresponds to the decrease of carbon species others than CO and $CO_2$. Oxidized hydrocarbons $C_xH_yO_z$ consumption could be enhanced with active species produced when ozone is destroyed or less produced when $E_d$ or temperature increase. O-atoms are available to produce active species (OH. $O_3$, O) which react with by-products to produce CO and $CO_2$.

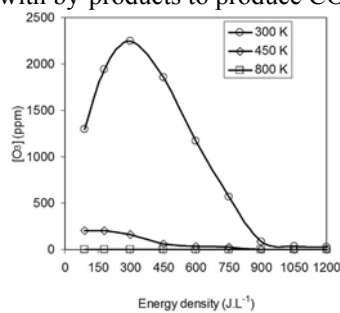

**Fig. 11.** Ozone concentration *vs* input energy density for different gas temperature

## 7. Conclusion

Chemical mechanisms in high voltage transient discharges (DBD) and in non thermal arc discharges (gliding, rotating and stationary discharges) lead to high production of active species. Depending on the inlet and on the nature of the discharge, the main oxidized species could be ozone, carbon monoxide and dioxide, nitrogen oxides, and $C_xH_yO_z$ species such as aldehyde, formic acid, ...

NOx, CO, $CO_2$, $H_2$, $C_2$-hydrocarbons are produced in non thermal arcs devices: Glidarc, Rotarc and stationary glow discharge Statarc. Because of high reactor outlet temperature, no ozone production is observed. Conversion of high hydrocarbon is available by using these reactors.

It could be supposed that main effect for decontamination and sterilisation would be involved by the NO production. NO in interaction with water can leads to acid mixture able to attack the organic matter.

Ozone can be easily produced in dielectric barrier discharges (DBD). Also DBD is able to produce various oxidized species. Maximum of $C_xH_yO_z$ production is obtained at "medium" values of energy density in wet air conditions. Adjustment of DBD reactor dedicated to oxidized oxygen species must be taking into account in order to optimize the reactive species production.

## References

[1] Penetrante B M and Schultheis SE (ed.) 1993 Non-thermal Plasma Techniques for Pollution Control (Part A and B) (New York: Springer-Verlag)

[2] Park JY, Tomicic I, Round GF, and Chang JS 1999 J. Phys. D: Appl. Phys. 32, pp 1006-1011
[3] Kim HH, Prieto G, Takashima K, Katsura S, and Mizuno A 2002 J. Electrostatics 55; pp 25-41
[4] Urishima K and Chang J S 2000 IEEE Trans. Dielect. & Elect. Insul. 7, pp 602-610
[5] Van Veldhuizen EM, Zhou M, and Rutgers WR 1998 Plasma Chem. Plasma Process. 18, pp 91-111
[6] Sun W, Pashaie B, Dhali SK, and Honea F I 1996 J. Appl. Phys. 79, pp 3438-3444
[7] Khacef A, Cormier JM, and Pouvesle JM, 2005 J. Adv. Oxid. Technol. 8(2), pp 150-157
[8] Khacef A, Cormier JM, and Pouvesle JM 2002, J. Phys. D: Appl. Phys. 35, pp 1491-1498
[9] Khacef A and Cormier JM 2006, J. Phys. D: Appl. Phys. 39, pp 1078-1083
[10] Khacef A, Pouvesle JM, and Cormier JM 2006, Eur.Phys. J.: Appl. Phys. 33, pp 195-198
[11] Gorce O, Jurado H, Thomas C, Djéga-Mariadassou G, Khacef A, Cormier JM, Pouvesle JM, Blanchard G, Calvo S, and Lendresse Y 2001, SAE paper N° 2001-01-3508
[12] Ouni F, Khacef A, and Cormier JM 2006, J. Chem. Eng. and Technol. 29(5), pp 604-609
[13] El Ahmar E, Met C, Aubry O, Khacef A, and Cormier JM 2006, Chem. Eng. J. 116, pp 13-18
[14] Aubry O, Met C, Khacef A, and Cormier JM 2005, Chem. Eng. J. 106(3), pp 241-247
[15] Rusu I, Cormier JM and Khacef A 2001, http://preprint.chemweb.com/ inorg-chem/0106001
[16] Masuda S and Nakao H 1990 IEEE Trans. Ind. Appl. 26, pp 374-383
[17] Dhali S K and Sardja I J 1991 Appl. Phys. 69, pp 6319-6324
[18] Penetrante B M, Hsiao M C, Bardsley J N, Merrit B T, Vogtlin G E, Wallman P H, Kuthi A, Burkhart C P, and Bayless J R 1996 Pure & Appl. Chem. 68, pp 1083-1087
[19] Evans D, Rosocha L A, Anderson GK, Coogan JJ, and Kushner MJ 1993 J. Appl. Phys. 74, pp 5378-5386
[20] Lesueur H, Czernichowski A and Chapelle A 1994 Int. J. Hydrogen Energy 19, pp 139-144
[21] Richard F, Cormier J.-M, Pellerin S and Chapelle J 1996 J. Appl. Phys. 79, pp 2245-2250
[22] Fridman A, Nester S, Kennedy LA, Saveliev A and Yardimci O M 1999 Prog. Energy Comb. Sci. 25, pp 221-231
[23] Frank N W 1995 Radiat. Phys. Chem. 45, pp 989-1002
[24] Cavvadias S, PhD Thesis 1987, University Paris 6 (France).
[25] McAdams RJ 2001, J. Phys. D: Appl. Phys., 34, pp 2810-2821.
[26] Chang CL, Bai H, Lu SJ 2005 Plas. Chem. Plas. Process., 25(6), pp 641-657.
[27] Motret O, Aubry O, Thuillier C, Lascaud M, Met M, Cormier JM, 2005 Proc. 17[th] Int. Symp. Plasma Chemistry, Toronto (Canada).
[28] Guo YF, Ye DQ, Chen KF, Tian YF 2006 Plas. Chem. Plas. Process. 26, pp 237-249.
[29] Kim HH 2004 Plasma Process. Polym. 1, pp 91-110.
[30] Blin-Simiand N, Jorand J, Belhadj-Miled Z, Pasquiers S, Postel C 2006, Proc. 5[th] ISNTPT, Oleron Island (France).
[31] Jarrige J, Vervisch P 2006 J. Appl. Phys. 99, pp 113303.1-113303.10.

## Keywords: